\documentclass[10pt, sigconf]{acmart}

\AtBeginDocument{%
  \providecommand\BibTeX{{%
    \normalfont B\kern-0.5em{\scshape i\kern-0.25em b}\kern-0.8em\TeX}}}

\usepackage[capitalise,noabbrev]{cleveref}
\usepackage{subfigure}
\usepackage[inline]{enumitem}
\usepackage[noend,ruled,linesnumbered]{algorithm2e}
\usepackage{graphicx}
\usepackage{float}
\usepackage{bm}
\usepackage{balance}
\usepackage{ifthen}

\definecolor{codegreen}{rgb}{0,0.6,0}
\definecolor{codegray}{rgb}{0.5,0.5,0.5}
\definecolor{codepurple}{rgb}{0.58,0,0.82}
\definecolor{backcolour}{rgb}{0.95,0.95,0.92}
\definecolor{ao(english)}{rgb}{0.0, 0.42, 0.24}
\definecolor{dark_green}{RGB}{10,100,70}

\usepackage{listings}
\newcommand{\systemname}{Rösti}
\usepackage{pifont}
\newcommand{\cmark}{\ding{51}}
\newcommand{\xmark}{\ding{55}}
\newcommand{\xstar}{\ding{87}}

\newboolean{submission}
\setboolean{submission}{true}

\usepackage{tikz}
\usetikzlibrary{shapes,arrows,positioning,calc,fit}

\ifthenelse{\boolean{submission}}{
    \newcommand{\toremove}[1]{ }
    \newcommand{\expected}[1]{ }
    \newcommand{\reviewer}[1]{ }
    \newcommand{\notes}[1]{ }

    \newcommand{\fixme}[1]{ }
    \newcommand{\todo}[1]{ }
    
}{
    \newcommand{\toremove}[1]{\color{red}{#1}\color{black}}
    \newcommand{\expected}[1]{expected: #1}
    \newcommand{\reviewer}[1]{\textcolor{blue}{[Reviewer: #1]}}
    \newcommand{\notes}[1]{\textcolor{ao(english)}{[NOTE: #1]}}
    
    \newcommand{\fixme}[1]{\color{red}{FIXME: #1}\color{black}}
    
    \newcommand{\todo}[1]{\color{blue}{TODO: #1}\color{black}}
    
}

\acmDOI{}
\acmConference[]{}{}

\begin{document}
\title{Declarative Concurrent Data Structures}

\author{Aunn Raza}
\authornote{These authors contributed equally}
\email{aunn.raza@epfl.ch}
\orcid{0000-0002-2586-3334}
\affiliation{%
  \institution{EPFL}
  \city{Lausanne}
  \country{Switzerland} 
}

\author{Hamish Nicholson}
\authornotemark[1]
\email{hamish.nicholson@epfl.ch}
\orcid{0000-0003-0306-3253}
\affiliation{%
  \institution{EPFL}
  \city{Lausanne}
  \country{Switzerland}
}

\author{Ioanna Tsakalidou}
\email{ioanna.tsakalidou@epfl.ch}
\orcid{0009-0001-4958-2270}
\affiliation{%
  \institution{EPFL}
  \city{Lausanne}
  \country{Switzerland}
}

\author{Anna Herlihy}
\email{anna.herlihy@epfl.ch}
\affiliation{%
  \institution{EPFL}
  \city{Lausanne}
  \country{Switzerland}
}

\author{Prathamesh Tagore}
\email{prathamesh.tagore@epfl.ch}
\affiliation{%
  \institution{EPFL}
  \city{Lausanne}
  \country{Switzerland}
}

\author{Anastasia Ailamaki}
\email{anastasia.ailamaki@epfl.ch}
\orcid{0000-0002-9949-3639}
\affiliation{%
\institution{EPFL}
    \city{Lausanne}
  \country{Switzerland}
}

\renewcommand{\shortauthors}{Raza, et al.}

\begin{abstract}




Implementing concurrent data structures is challenging and requires a deep understanding of concurrency concepts, synchronization techniques, and careful design to ensure correctness, performance, and scalability. 
Further, composing operations on two or more concurrent data structures often requires a synchronization wrapper to ensure the operations are applied together atomically, resulting in serialization and, thereby, giving up the performance benefit of the individual data structures.
%
Database management systems (DBMS) guarantee atomicity, consistency, isolation, and durability (ACID), providing a generalized implementation of concurrency control over user data. 
Given these properties, a DBMS would seem to be a good fit for implementing concurrent data structures. 
However, DBMSs are over-generalized for this use case, which results in a failure to match the performance of a specialized data structure implementation.


In this paper, we make the case for Declarative Concurrent Data Structures (DCDS), a framework for the automatic generation of concurrent data structures from a serial specification.  
In DCDS, users declare the attributes and methods needed for their desired data structure at design time through an embedded domain-specific language (DSL). 
At build-time, DCDS automatically injects concurrency control protocol, generating a concurrent intermediate representation (IR) which is compiled to machine code. 
A declarative interface for designing data structure enables efficient composability through co-optimization of component structures; optimizations are applied to both the composed serial specification and then the generated concurrent IR. 
We realize the DCDS framework in our prototype system \systemname{} and experimentally show that data structures declared in \systemname{} can be efficiently composed through co-optimization of both their logical functionality and the generated concurrency control protocol.
Our evaluation shows that composing a map and a list to create a recency-sorted container can benefit up to 2X performance scalability in \systemname{} compared to an open-source library implementation.
%
We demonstrate the applicability of DCDS as an in-process OLTP by comparing it with in-memory DBMS, Proteus, and showing up to 2X performance gains.

\end{abstract}

\settopmatter{printfolios=true}
\settopmatter{printacmref=false}

\maketitle



\section{Introduction}
\label{sec:intro}






Modern hardware offers increasingly high levels of parallelism; a current generation commodity server can have up to 512 hardware threads (256 physical cores) with two sockets~\cite{amd_epyc_genoa_cloud}. 
In order to correctly and efficiently utilize this hardware, developers are expected to write scalable concurrent code that executes operations in parallel.
Writing performant and correct concurrent programs is notoriously challenging and error-prone~\cite{DBLP:conf/asplos/LuPSZ08}. 
In the context of this paper, correctness entails ensuring no race conditions, live- or dead-locks, and that conflicting concurrent operations do not produce inconsistent results.

Concurrent programming requires a deep understanding of concurrency concepts (thread model, critical sections, memory consistency models, etc.), synchronization techniques (locks and latches, lock-free, read-copy-update paradigm, etc.), and careful design considerations to ensure correctness, performance, and scalability of the data structure. 
Developing a concurrent (thread-safe) data structure implementation is vastly more time-intensive than the non-thread-safe equivalent due to the myriad of bugs that concurrency enables~\cite{drdobbsLockFreeCode, DBLP:books/daglib/0020056}. 

Most applications and developers rely on highly optimized libraries for concurrent data structures, like Intel's oneAPI TBB~\cite{reinders_intel_2007,githubGitHubOneapisrconeTBB}, Meta's Folly~\cite{githubGitHubFacebookfolly}, Boost~\cite{githubGitHubBoostorgboost}, etc., or use simple naive concurrency guards for non-performance critical code.
Essentially, developers often have to make a decision to either reuse an existing available implementation in the form of a library or implement one from scratch. 
Implementing concurrent data structures from scratch requires a large time investment. In contrast, library data structures are designed for general use and thus often have superfluous features for a specific case or are missing niche application-specific functionality or performance optimization.

While data structure libraries promote code reusability, it is rare that an implementation of the ideal concurrent data structure for a particular use case already exists. In what follows, we explain the challenges and trade-offs associated with reusing existing specialized concurrent implementations that are battle-tested and performant in the use cases they were designed for but may or may not perform adequately in other use cases:

\textbf{Functionality-performance trade-off.}
The implementation of a concurrent data structure often either aims to support functionality specialized for a particular use case or to be general-purpose and reusable, offering a broad range of functionality.
In the former case, the data structure is specialized to the target requirements and, therefore, either becomes unusable or merely provides functionality but not performance for the general case.
For example, a linked list can be specialized for insertions and removal operations only at the head and tail of the list and therefore only requires concurrency control on the head and tail pointers, while if there is a need for erasing or updating specific nodes in a list, the developer may either optimize the concurrency control granularity to be on individual nodes (such as hand-over-hand locking~\cite{DBLP:journals/acta/BayerS77}) or merely provide functionality by locking both, head and tail pointers, simulating a full data structure lock.
In the case where generalization and reusability are the primary goals, a developer might create a functionality to retrieve the size of the linked list, that is, the number of elements in the list. To implement this functionality, the data structure will either maintain a size variable, and hence the size variable becomes a concurrency bottleneck for other operations given that it needs to be updated on all insertions and removals, or will iterate the entire list to calculate the size, which again requires concurrency control on the full data structure.
In both cases, functionality comes at a cost in concurrent data structures, and thereby, implementations either trade performance for functionality, hence generalization, or vice versa, hence specialization to specific use cases.

\textbf{Composing concurrent data structures}
Composability is at the core of application development.
Composing concurrent data structures requires the atomicity of multiple operations across different data structures.
However, composing concurrent data structures is not trivial. It often leads to serializing operations across all data structures, giving up on the optimization made for an individual concurrent data structure's scalability.
For example, consider implementing a least-recently-used (LRU) list through a hash table for existence checking and a linked list for eviction ordering.
An insert operation requires adding a node to the linked list and the hash table to be atomic.
Even if the hash table and linked list are individually concurrent and scalable, to have an atomic operation across both, a synchronization wrapper will cause the operations to be executed serially with a lock guard, hence losing the benefits from individual concurrent and scalable implementations.

We aim to enable developers to create scalable concurrent data structures tailored to their use case with the same effort as designing a non-thread-safe data structure. 
To achieve this, we propose a framework for the automatic generation of concurrent data structures from a serial specification. 
Users create a sequential specification of the functionality of their data structure through an embedded domain-specific language (DSL), declaring the \textit{what}, and the DSL compiler can generate an optimized thread-safe implementation.
The DSL makes this possible by clearly defining the entry points to data structure methods from the host language and by restricting side effects to only operations that mutate data structure attributes. 
Together, these DSL features enable precise and complete determination of the read and write sets of data structure methods at compile time. 
A user can compose data structure specifications to create a new data structure specification; a data structure may have attributes that are other previously specified data structures, and the DSL compiler will co-optimize them at build time. 
The host language can access the generated data structure either through library import or through generating and linking the code in the application process using the data structure.




In summary, we make the following \textbf{contributions:}
\begin{itemize}
    \item We make the case for declarative concurrent data structures (DCDS). DCDS enables developers to design and declare their data structures in a functionality-oriented serial manner, and then, at build-time, DCDS optimizes and generates concurrent data structures.

    \item We implement the DCDS framework in \systemname{}, which uses the LLVM code-generation framework to compile a DCDS specification into an executable concurrent data structure library.

    %
    \item We show that data structures declared in \systemname{} can be efficiently composed by co-optimizing their logical functionality and the generated concurrency control protocol. 
    Our evaluation shows that composing a map and a list to create a recency-sorted container can benefit up to 2X performance scalability in \systemname{} compared to open-source library alternative and show use case driven automatic optimization of doubly to singly linked list in case use case of a FIFO list.
    

    \item We demonstrate the general applicability of DCDS and \systemname{} by comparing it with an in-memory DBMS, Proteus~\cite{proteusweb}; on the YCSB benchmark, we show up to 2X gains as in-process transaction processing.

\end{itemize}

DCDS is the first step towards declarative system development, co-optimizing composite data structures based on the functionality and usage requirements, specializing concurrent implementation to the actual requirements at build time while providing generalization during development time.

\section{Background}
\label{sec:making_case}



\begin{table*}[ht]
\centering
\caption{A summary of common techniques for implementing concurrent data structures}
\label{tab:approach_comparison}
\begin{tabular}{@{}lrlll@{}}
\toprule
\multicolumn{1}{c}{Method}                                                                         & Ease of development  & Composability &  Performance        \\ \midrule
Coarse-grained locking                                                                             & \xstar\xstar\xstar\xstar     & \cmark     & \xstar             \\
\begin{tabular}[c]{@{}l@{}}Hand optimized \\ (e.g. fine-grained locking or lock-free)\end{tabular} & \xstar                       & \xmark     & \xstar\xstar\xstar\xstar \\
Software Transactional Memory                                                                      & \xstar\xstar                 & \cmark     & \xstar\xstar       \\
DCDS                                                                                               & \xstar\xstar\xstar           & \cmark     & \xstar\xstar\xstar       \\ \bottomrule
\end{tabular}
\end{table*}


Concurrent data structures and algorithms enable parallelism and, thereby, scalability.
Amdahl's law~\cite{DBLP:journals/computer/HillM08,DBLP:conf/afips/Amdahl67} states that the overall speed up of a parallelized task is limited by the proportion of the task that cannot be parallelized; this is often referred to as the scalability or concurrency bottleneck.
Concurrent data structures guarantee semantic correctness when operations on the data structure are invoked concurrently, that is, more than one overlapping invocation, thus eliminating race conditions~\cite{DBLP:journals/loplas/NetzerM92}.

The methods used to implement concurrent data structures can be categorized into a spectrum, ranging from offloading to a generalized external system, such as a database management system (DBMS), to hand-optimized and tailored implementations. 
In this section, we lay out the techniques for constructing concurrent data structures and discuss their benefits and drawbacks.
We classify different approaches for developing a concurrent data structure based mainly on three axes, ease of development, composability, and performance.
\cref{tab:approach_comparison} summarizes the differences between each representative approach in the spectrum. 

\textbf{Manually Implemented Data Structures.}
Coarse-grained locking, using a lock to guard an entire data structure to ensure mutual exclusion, is the simplest technique used to construct a concurrent data structure (CDS), but its simplicity often comes at the cost of performance.
Although such a data structure is thread-safe by definition, it does not provide performance scalability as it serializes all concurrent operations through the lock guard.
At the other end of the spectrum, hand-optimized fine-grained lock-based and lock-free approaches use more granular synchronization to achieve excellent performance in highly concurrent workloads.
However, it is extremely challenging to implement both performant and correct data structures with these methods~\cite{drdobbsLockFreeCode, DBLP:books/daglib/0020056} even with improved thread-safety tooling~\cite{DBLP:conf/scam/HutchinsBS14, DBLP:conf/pldi/QinCYSZ20}. 
As a result, application-specific data structures are rarely built from scratch with these techniques except in the most performance-critical components of a system.
Most general-purpose library CDSs use lock-free or fine-grained locking, but a library CDS will not be optimized for a particular application's requirements. 
Further, lock-free and fine-grained locking CDSs lack mechanisms to atomically and performantly compose operations across multiple data structures~\cite{DBLP:journals/cacm/HarrisMJH08}. 
These contrasting methods serve to highlight the trade-offs facing developers: ease of development and poor performance under contention versus high development complexity and high performance.

\textbf{Database Management System.}
DBMSs are generalized data management systems for multi-user, multi-tenancy environments and are designed to handle concurrent data access, guaranteeing ACID (atomicity, consistency, integrity, and durability) properties.
To provide these properties, DBMSs implement sophisticated concurrency control (CC) mechanisms to synchronize concurrent accesses.
DBMSs provide a declarative interface to define the data structure and the operations over it, most commonly through SQL. 
DBMSs can specialize for the workload, as the declarative interface allows the DBMS to choose how to execute the query. 
However, this adaption only happens at runtime and with limited information on how other transactions may operate over the managed data. 
In order to preserve correctness, the DBMS can only conservatively apply optimizations.

A DBMS introduces unnecessary overheads for implementing a CDS due to their generality and additional layers of abstraction.
For example, a runtime query optimizer is necessary for ad hoc transactions, but if the functionality of the data structure is known ahead of time, a runtime query optimizer is not necessary.
Further, features such as durability and multi-tenancy are also, in general, not necessary for an application data structure. 
Finally, the application code must communicate with the DBMS. 
Most DBMS follow the client-server model, and thus, the application must serialize requests, communicate over a protocol such as JDBC or ODBC, and deserialize responses, all adding substantial overhead compared to an in-process data structure. 
While some of these costs may be mitigated using an embedded DBMS, an embedded DBMS is still limited in the optimizations it can apply to the operations over the data structure.  
Utilizing a DBMS to implement a CDS can simplify the development process and provide composability and ACID guarantees through the declarative interface.
However, this approach comes with performance overheads for a CDS due to unnecessary DBMS features and the DBMS being constrained in the optimizations it can apply as it cannot know the full set of operations over the data structure.

\textbf{Software Transactional Memory.}
Software transactional memory (STM) provides declarative concurrency by moving the intricacies of currency control behind a transactional interface.
Like DBMSs, this separation of concerns alleviates some of the complexity of implementing CDSs. 
Further, separate data structures can be atomically and consistently modified under the STM model as part of a single transaction.

\fixme{maybe distinguish between auto STM injection/code-generation and explicit STM. Low priority}
However, STM has its own challenges; it employs a generalized CC algorithm without application-level knowledge. 
This limits the ability to apply workload-specific optimizations, obliging STM to adopt a conservative approach to concurrency that may fall short of yielding optimal performance.
Developers must familiarize themselves with transactional semantics to use STM effectively.
Identifying which parts of the code should be transactional and at what granularity requires a nuanced understanding of both the data structure and the concurrency semantics to achieve both correctness and performance~\cite{DBLP:journals/cacm/CascavalBMCWCC08}.

\textbf{Declarative Concurrency Control.}
An implementation of a CDS must effectively implement a variant of a concurrency control (CC) algorithm. 
The implementation of the CC algorithm defines which operations on the data structure are atomic, which may be altered and tailored to fit specific needs.
CC algorithms have been standardized and are implemented in all transactional database management systems (DBMS) as well as in STM systems.
Both DBMS and STM provide a separation of concerns between the semantics of an operation and the thread-safe execution of the function.  
Different CC protocols~\cite{DBLP:journals/pvldb/WuALXP17} optimize for different workload characteristics (read-heavy, write-heavy, etc).
The guarantees and goals of the CC mechanism are invariant with the specific implementation and workload. 
They enable atomicity across operations while allowing read/write operations to be interleaved across attributes.
DBMS and STM both aim to enable declarative concurrency control but trade performance for ease of development or vice versa.

\fixme{Hamish: Someone who is not me give an opinion on the table}

\textbf{Declarative Concurrent Data Structures (DCDS).}
DCDS converges the gap between ease of use and performance. 
While hand-tuned approaches can leverage maximum application-level knowledge, they are error-prone and time-consuming to build.
In contrast, DBMS and STM offer declarative interfaces, abstracting away the concurrency control mechanisms.
However, their limited application-level knowledge restricts the optimizations they can make.
DCDS bridges this gap by providing a framework that allows developers to specify their data structures declaratively and transparently convey application-level knowledge across the abstraction boundary. This enables the system to optimize these structures for the specific application and workload at build time.
This approach combines the ease of development of declarative systems with the performance advantages of hand-tuned data structures.

\section{DCDS Manifesto}
\label{sec:dcds_overview}
\tikzset{
block/.style = {draw, fill=white, rectangle, minimum height=3em, minimum width=1.5cm, align=center},
tmp/.style  = {coordinate}, 
sum/.style= {draw, fill=white, circle, node distance=1cm},
input/.style = {coordinate},
output/.style= {coordinate},
pinstyle/.style = {pin edge={to-,thin,black}
}
}


\begin{figure}[!htb]
\centering
\begin{tikzpicture}[auto, node distance=2cm,>=latex']
    \coordinate (center) at (0,0);
    \node [block, xshift=-0.1cm](frontend){frontend DSL};  %
    \node[draw,circle,minimum size=1.3cm,inner sep=0pt, below= 0.4cm of frontend](IR) {DCDS-IR};
    \begin{scope}[below of=IR, name=backend_scope, yshift=-1.3cm, xshift=-1cm]
            \node [block] (logical_opt) {Logical \\ Optimizer};
            \node [block, below= 0.2cm of logical_opt] (physical_opt) {Physical \\ Optimizer};
            \node [block, right= 0.3cm of logical_opt] (concurrentization) {CC \\ Injector};
            \node [block, below= 0.2cm of concurrentization] (compiler) {Code \\ Generator};
            \node [draw,  fit=(logical_opt) (physical_opt) (concurrentization)  (compiler),inner sep=0.15cm, label=below:Backend] (backend){};
    \end{scope}
    \node [block, right= 0.5cm of backend, label=below:Generated Datastructure](generated_code){\texttt{MyDS.so}/\texttt{MyDs.hpp}}; 
    \draw [draw,->] (frontend) -- (IR);
    \draw [draw,->] (IR) -- (backend);
    \draw [draw,->] (backend) -- (generated_code);
    \end{tikzpicture}
\caption{The DCDS framework. } 
\label{fig:dcds_framework_block}
\end{figure}
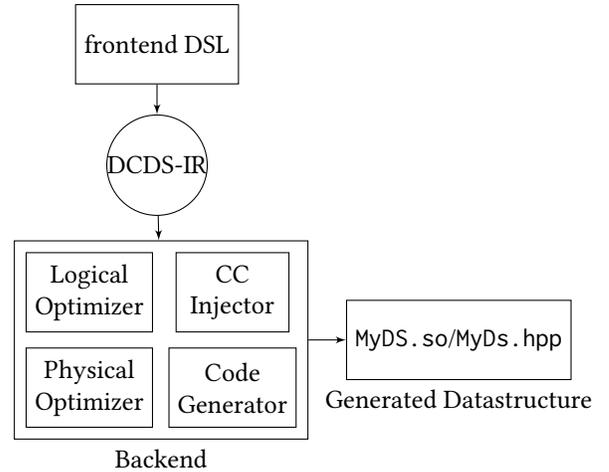

The DCDS framework takes inspiration from the declarative interfaces of DBMSs.
In the DCDS framework, users declare their data structure's attributes and methods as though each method call were to be executed serially, allowing the user to customize the functionality of the data structure to their needs. 
This means that DCDS knows the semantics of all methods that can modify the state of the data structure, which enables DCDS to co-optimize the generated thread-safe implementation of each of the methods.

\Cref{fig:dcds_framework_block} shows the logical components of the DCDS framework. 
A user-facing frontend, such as a C++ API, is used to generate an intermediate representation (DCDS-IR) of a data structure. 
The complete DCDS-IR output of the frontend for a particular data structure is a \textit{data structure specification}.
DCDS-IR is reusable, shareable, and composable. 
Developers can declare a new data structure from scratch, reuse an existing specification, or compose multiple data structure specifications together to form a new specification.
The data structure specification is designed to be easily modifiable to customize the resulting data structure to the application's needs; the attributes and methods of an existing specification can be added, removed, or replaced.  

The backend consumes the data structure specification in DCDS-IR and generates the data structure in a format compatible with the user's application, for example, a shared object and a header file for C/C++ applications. 
The backend consists of four stages. 
First, the \textit{Logical Optimizer} applies semantic optimizations over the serial specification of the data structure,  \fixme{for example.... which makes CC injection more effective ...}.
Then, the \textit{CC Injector} applies a pass to the DCDS-IR representation of the data structure to create the thread-safe DCDS-IR form of the data structure. 
The specific CC algorithm used is not prescribed by the framework and is a choice left to the implementation of the framework. 
The \textit{Physical Optimizer} applies a second round of optimization passes over the now thread-safe representation.
This stage is closely tied to the choice of CC algorithm, as any optimizations made must preserve the guarantees of the CC algorithm. 
Finally, the \textit{Code Generator} lowers the thread-safe and optimized DCDS-IR into an executable format and generates any source files necessary for integrating with the target languages. 
The Logical Optimizer and Physical Optimizer only improve the performance of the resulting CDS; they are not necessary for correctness. 
For example, a naive backend, but one that still generates a correct CDS, could omit optimization and use a coarse-grained lock for each data structure. 
\fixme{Forward ref to next section where we detail our design/implementation?}

The design of DCDS framework is based on principles of clean separation of concerns and abstractions: The user defines the functionality, that is, declaring the \textit{'what'}, while DCDS transparently optimizes and generates a thread-safe CDS as an output.
The user does not need to know the underlying CC mechanism used in order to create a correct and performant DCDS. 
The key is the unobtrusive transfer of application knowledge to the backend through the serial data structure specification, in particular, the specification of the public methods exposed by each data structure.
As the specification declares only the functionality of the methods, and these methods are the only code that will modify the state of the attributes, DCDS enables the co-optimization of the implementation and CC injection for all the declared methods.

\fixme{TBD: in next submission.}
\fixme{is this the correct spot for writing about scoped optimization which a compiler cannot do.}
\fixme{Hamish: yes, will try as an extension of paragraph, new paragraph is hard since the ideas are same/similar}

\section{\systemname{} System Design: DCDS for C++}
\label{sec:system}





\begin{figure*}[ht]
    \centering
    \includegraphics[width=\textwidth]{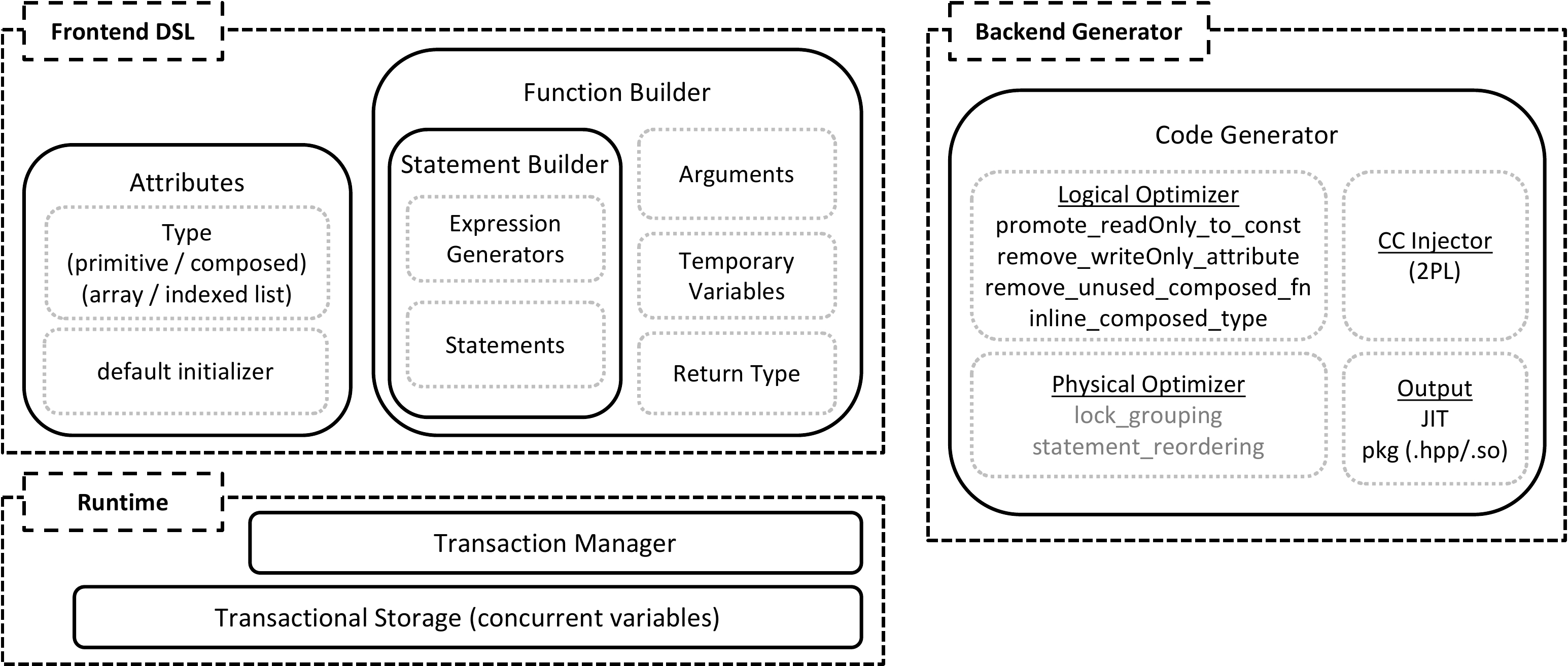}
    \caption{\systemname{} System Architecture}
    \label{fig:system_overview}
\end{figure*}


In this section, we provide the system design of \systemname{}. \systemname{} is an implementation of the DCDS framework in the C++ programming language and enables users to generate CDS through its declarative interface, an embedded DSL. 
Figure~\ref{fig:system_overview} shows the overall representative system architecture of the \systemname{}. In what follows, we first explain the different components of the system, including the declarative interface which interacts and builds the DCDS IR, and then explain the IR optimization phases, concurrency control injection, storage layer, and the methods to export and utilize the generated CDS.

\todo{somewhere in the case, we need to say that we chose DSL over annotations to limit program analysis scope, which is super-complex IMO.}
\todo{Hamish: \^ will be resolved with compiler comparison / scoped optimization paragraph in preceding section}

\subsection{Frontend DSL}

\systemname{} implements the DCDS framework and provides a declarative interface through an embedded domain-specific language (DSL) in C++ programming language. 
%
%
\systemname{}'s interface provides a subset of functionality that a generic programming language provides, enabling developers to declare serial specifications of the required data structures through the embedded DSL in C++. 
The embedded DSL includes constructs and abstractions for attribute and function declarations in a typed manner, statements and expressions for defining a function body, and context for declaring workload- and application-specific requirements.
%
In the following, we explain the details of our interface and how developers can utilize it to declare serial data structures and generate CDS.

\textbf{Builder.} 
The builder is the entry build to the \systemname{} DSL.
It encapsulates and provides interfaces for declaring, composing, and building data structure specifications.
All the data structure attributes and functions are declared through the builder interface. For composing one or more data structure specifications, a composed type can be either created through the builder instance or one builder instance can be registered as a composed type with other builder instances.

\textbf{Attributes.} The attributes of a data structure are declared through the builder.
These attributes are assumed to be concurrent variables and would be protected through the CC mechanism at runtime.
Developers can declare attributes through the builder by specifying the type of the attribute and an initial or default value.
In \systemname{}, the types of the attribute can be either one of the primitive types, that is, 8-, 16-, 32- and 64-bit integers, 64-bit double, fixed-length null-terminated strings \footnote{We have not yet implemented support for variable-length strings}, or, can be a composite type, that is, another builder instance, essentially, composing other data structures built through \systemname{}.
Further, \systemname{} also allows two additional complex types, fixed-length arrays and a key-indexed map. Fixed-length arrays pre-allocate the required storage at instance initialization time, whereas the key-indexed map is implemented through a secondary index and allocates records on demand. Having array and index types at the interface layer enables \systemname{} to optimize and specialize the granularity of concurrency control based on the functional requirements at build-time.

\systemname{} exposes two means for composing data structures at the interface layer. The first is by declaring a (logical) pointer, this enables linked-list style data structures that store a mutable pointer to a data structure of the composed type.
The second is by creating an attribute in the main data structure of the composed data structure's type. 
The main difference between the two methods of composability is that pointers are mutable, and thus require concurrency control to dereference the pointer before methods of the composed type can be called. 
In contrast, accessing a composed data structure that is an attribute within the primary data structure does not require concurrency control as, logically, no pointer is dereferenced. In both cases, when methods of the composed type are called there will be CC to protect internal mutation of the composed data type.
Internally, composite attributes are implemented through physical record references, similar to a foreign-key reference in a DBMS, though this detail is not exposed in the interface.
We explain the physical storage of the primitive and composite attributes of a data structure in \cref{sec:storage_layer}.
This separates the the physical instances of each data structure in the storage layer, while transparently composing the contained data structure.

\textbf{Expressions Builder.}
The implementation of a data structure requires manipulating and transforming the data itself.
In \systemname{}, this is done through expressions, which are exposed through the expressions builder, similar to one implemented by DBMS~\cite{DBLP:journals/pvldb/KarpathiotakisA16}. 
Expressions mimic the projection operation from relational algebra, operate on one or two input variables, and return a transformed result variable.
Examples of expressions include \textit{ADD}, \textit{SUM}, \textit{SUBTRACT}, \textit{isNullPtr}, etc.. For simplicity reasons, expressions can only take variables or compile-time constants as input and generate output temporary variables, essentially, expressions are transformations without side-effects in \systemname{}. This limitation is technical and only simplifies the development of \systemname{} and can be easily extended to operate on attributes also by implicitly creating intermediate temporary variables.

\textbf{Function Builder.}
The function builder provides the interface for declaring the data structure's methods, that is, its functionality.
A function builder is created through a builder instance and is implicitly registered with the corresponding data structure. Declaring function requires the name of the function, return type, and optional arguments, if any. Function arguments can be either a value or a pointer to one of the primitive types in \systemname{}. 
Further, the function builder allows the declaration of typed temporary variables within the function body with the function-level scope. Temporary variables are similar to attributes but are contained within a function body and, therefore, do not require synchronized access across concurrent invocations.
In what follows, we refer to temporary variables and function arguments interchangeably as variables except when explicitly noted, and we use the term attribute specifically for data structure's attributes.

\textbf{Statement Builder.}
The statement builder provides the interface for declaring operations in order within a function body. \systemname{} implements the following types of statements and, by design, can be extended in order to incorporate more features and mimic a generic programming language while restricting operations that may cause side effects, thereby hindering CC injection, specifically in terms of rollback in case of a transactional abort.

The first two types of statements are read and update statements which directly interact with the data structure attributes.
A \textit{read statement} takes a source data structure attribute and a temporary variable as a destination.
A \textit{update statement} takes a target data structure attribute and either a temporary variable or a function argument as the data source. For collection types, fixed-length arrays, or key-indexed maps, the read and update statement takes an additional index or key parameter, respectively. Additionally, a key-indexed map provides contains and insert functionality.
Both, the read and update statement modifies the data structure state and will be protected by the CC mechanism.
%

Then, \systemname{} provides an interface to build \textit{conditional statements} for creating branches;
It takes an expression as input which evaluates to a boolean, and it returns a pair of empty lists of statements.
The lists of statements are the if and else branches of the conditional statement. \systemname{} enforces having at least one statement in the if branch, while the else branch can be empty. Further, conditional statements can be nested, similar to control-flow-based programming language.
We do not provide an explicit merge branch for statements that follow the execution of the conditional statement and one of its branches. 
This is in contrast to LLVM-IR which requires an explicit merge block.
Any statement added to the function body after the conditional statement will be executed after the conditional statement and the list of statements of the if or else branches.

\systemname{} also provides \textit{create}, \textit{delete} and \textit{method-call} statements for the composed types. 
A create statement takes a composed data structure type as input and returns a temporary value that stores a reference to a new instance of the composed data structure.
These functions enable the creation and use of composed types within the context of primary data structure.
For example, a linked list is a composition of the node type.
In the insert method of the linked list, the user would first create a new node, call a method of the node to update the node's next attribute to point to the current head of the linked list, and then set the linked list head attribute to store the reference to the newly created node.
The initialization of a composed type differs from the initialization of the primary data structure; the methods of the composed type are not automatically a part of the data structure interface that the compiled primary data structure will expose to the user. 
Internally, this enables optimizing the construction of the composed type within the callee method of the primary data structure. 

Finally, \systemname{} provides the functionality to declare \textit{return statements}, which are used to return the control flow from the function itself. Return statements can be either of type void or can return a variable declared within the function scope. \systemname{} requires explicit return statements.




\subsection{Backend Generator}

\systemname{}'s backend generator implements optimizer passes and utilizes LLVM to generate code for the CDS given the declared serial specification through the frontend DSL. 
First, the logical optimizer operates on the \systemname{}-IR to optimize the serial functionality, including but not limited to applying optimization passes for dead-code elimination and inlining.
Then, the backend injects a CC protocol to enable thread-safe data structure operations. 
The \systemname{} design includes a physical optimizer but is left for future work in the scope of this paper.
The physical optimizer will operate on concurrent IR and optimize the injected CC operations; For example, by lock grouping across attributes and reordering lock acquisition statements to reduce the runtime cost of aborts. Lastly, the concurrent IR is lowered to machine code and a C++ wrapper for the generated code is created.
%

\textbf{Logical Optimization.}
\systemname{} logically optimizes \systemname{}-IR similarly to compiler-passes or logical query optimization in a DBMS. 
\systemname{}-IR enables many optimizations, which a general compiler cannot do given the lack of information and pessimistic proof of results. 
Further, in contrast to DBMS, \systemname{} has complete knowledge of the workload and by construction, ensures that no other operation will be performed on the declared data structure. This enables \systemname{} to optimize and eliminate any unused attribute or functionality. 
Currently, the main logical optimizations in \systemname{} are targeted towards eliminating dead and redundant code.

All of the entry points, the methods that can be called from C++ user code, that can mutate a data structure are within the optimization scope of \systemname{}-IR. This enables tracking of the complete read and write sets for all functions in the data structure at compile time. 
Declared attributes are private to a data structure and can only be accessed through the declared functions.
Similarly, any functionality of the composed types can only be used by the contained type.
This allows \systemname{} to aggressively optimize composed types and attributes.
Initially, and after each optimization pass, the optimizer recursively creates read and write sets for all the functions in the data structure to maintain accurate attribute usage information across the top-level and composed data structures. 
In the following, we explain the different optimization passes that enable \systemname{} to remove redundant or dead code and eliminate overheads due to extraneous synchronization.

1) \textit{Remove unused functions from composed types:} \systemname{} traverses the DCDS IR and then removes all unused functions of the composed type. \systemname{} cannot detect the usage count of the exposed, that is, public functions of the main top-level data structure, but can detect the usage of the functionality of composed types. Recursively removing unused functions of composed types leads to the removal of unnecessary synchronization and unused attributes that are not required for the functionality of the main data structure. 

2) \textit{Remove unused attributes:} \systemname{} removes all attributes that are neither read nor written in any declared functions. Removal of unused attributes is performed recursively, on the primary data structure and then composed types.

3) \textit{Convert read-only attributes to constant expressions:} \systemname{} also detects if a data structure attribute is just read-only, and for each read-only attribute, it converts the attribute to a constant expression for compile-time constants.

4) \textit{Removing write-only attributes:} \systemname{} detects and removes write-only attributes as well as the corresponding update statement. This enables \systemname{} to remove unnecessary synchronization overheads for write-write conflicts that have no observable side effects. 
This is in contrast to a DBMS, which cannot remove any attribute from the declared schema, as a DBMS cannot guarantee that the target attribute will not be referenced again in an unknown future workload.


\textbf{CC Injector.}
\systemname{} injects concurrency control primitives in the DCDS IR after performing the logical optimizations phase. Currently, \systemname{} injects strict two-phase locking (S2PL)~\cite{DBLP:books/aw/BernsteinHG87} CC with NO\_WAIT deadlock avoidance protocol.
DCDS CC injection and optimization differs from a general DBMS as DCDS has complete information on concurrent variables and their conflicts within the critical sections of the data structure's methods.

CC injection proceeds by visiting each function, which marks the transactional scope, and then each statement in that function. For each read or write statement encountered that operates on a data structure attribute, a shared or exclusive lock statement is injected before the read or the write statement, respectively, if the lock is not already acquired in the current transactional scope. If the CC injector encounters a request for an exclusive lock while a shared lock already exists, it converts the existing shared lock into an exclusive lock. 
As per the strict two-phase locking protocol, the lock release statement is added at the very end of the function, before the return paths, to prevent cascaded aborts.


In the case of a method-call statement, a call a function of a composed type, the CC injector first checks if the referenced attribute is \textit{nascent}, i.e., it was created within this scope and has not been written to a shared location, essentially denoting side-effect free until that point. 
If the referenced attribute is not nascent, the injector inserts a lock statement prior to the corresponding method call statement. This optimization prevents extraneous locks in the critical path. For example, in a push method of a linked list, the method creates a new node, sets the payload and the next pointer to the current head, and then changes the head pointer to this node. In this example, the new node is nascent until it is written as the head in the linked list data structure.
The type of lock, that is, shared or exclusive, depends on whether the function being called is a \textit{const} function or not.
A const function in DCDS is similar to a const function in C++; however, in DCDS, const functions are deduced rather than requiring an annotation from the programmer. 
Detecting if the function is const or not is done during the IR construction phase, where if the function does not modify any attribute of the contained data structure, it is marked as a const function.




\textbf{Exporting \systemname{}-generated CDS}
\systemname{} supports two methods of exporting the generated data structure, library export, and generating code within the process of the application that will use the data structure.

\textit{1) Library export: }
In library export mode, \systemname{} generates a pair of .hpp (header) and .so (shared object) files for each data structure, that can then be linked with any other C++ application or system. Once included and linked, the data structures included in the exported header can be used like any other imported data structure or library in the C++ programming language.

\textit{2) In-process code-generation: }
In integrated applications where \systemname{} is included as a library, \systemname{} supports generating code and linking the code to the application at the runtime of the application. In this mode, the developer can declare the data structure specification and call the $jit\_compile\_and\_load()$ function, which specializes and then builds the data structure through LLVM JIT code-generation infrastructure. This dynamically links the compiled data structure to the application and loads its symbols. On data structure instantiation, \systemname{} calls the constructor of the corresponding data structure and wraps the data structure in a $jit\_container$ type which contains the CDS method's addresses and return types.
JIT mode enables quick prototyping and testing of expected functionality, and also enables runtime creation and manipulation of data structure. 

\subsection{Runtime}
\label{sec:storage_layer}

CDS generated by \systemname{} instantiates two auxiliary data structures in the application runtime; transaction manager and transactional storage. Runtime data structures are required to enable concurrent transaction processing and, thereby, concurrent execution of data structure operations. In the following, we explain the details of the required auxiliary data structures in the application's runtime.

\textbf{Transaction Manager.} 
The transaction manager handles concurrent transactions for the \systemname{}-generated CDS. 
Transactional namespaces partition transaction managers. 
Each data structure is associated with a single transaction manager at runtime, and to partition transaction management across data structures, a developer may supply an optional namespace or use a default namespace during data structure instantiation, which gets or creates a new transaction manager if it does not exist already.

Each call to a data structure method from C++ user code data structure operation is a transactional scope. 
At the start of this method call, CDS calls begin\_txn on the assigned transaction manager and receive a transaction object. 
The transaction object stores the transaction status, the log, and a list of acquired shared and exclusive locks. 
Each data mutation operation with observable side effects is stored in the log object, which in turn may be used to roll back a transaction in the case of an abort. 
In S2PL CC with NO\_WAIT, the data structure operation can call end\_txn to either successfully commit at the end of the transaction block or abort and rollback where it fails to acquire a lock on resource. 
In \systemname{}, there are no user-generated aborts, but only when the CDS operation cannot proceed given the underlying CC protocol.

\textbf{Transactional Storage.}
\systemname{} stores the backing data for each data structure in a table, each instance of the same data structure type is one row in a table. This is conceptually similar to storing relational tables in a row format.  
The storage layer comprises two data structures: the table registry and the tables. The table registry is implemented as a runtime singleton and keeps track of all the tables currently active in the application runtime.
Currently, each data structure is backed by a single table in a single transaction namespace (table names are prefixed with the namespace to partition storage across transactional namespaces), where each column denotes the attributes of the data structure.
We do not perform column splitting at this stage and leave it for future work.
A secondary table is created for fixed-sized arrays and key-indexed maps. Currently, the storage layer is not NUMA-aware; hence, it does not perform any partitioning, and the memory location from the allocation is entirely dependent on the allocator's preferences. 
Each table stores the full record size, attribute types, and their size for allocation and deallocation purposes, and upon insert or delete requests, allocates memory from the system's memory allocator. In our implementation, we use OneAPI TBB's memory allocator~\cite{githubGitHubOneapisrconeTBB}.
Further, each row is associated with metadata containing the unique row identifier and the locks and latches for the corresponding row.

\systemname{} generates the constructor of each data structure at build time. 
It first checks if the tables required for this data structure are already created, and if not, generates a call to initialize the tables. 
This process only happens for the first creation of the first instance of each data structure type.   
Then, for the primary data structure and each composed data structure, a record is allocated in the corresponding table, and the attributes in the record are initialized with the default values.
Inside of \systemname{}, records are identified by a 64-bit \textit{physical record reference}. 
The top 16 bits represent the $table\_id$, while the remaining 48 bits store the offset of the record in the table.
For fixed-size arrays and key-indexed maps, a secondary table is created. Fixed-sized arrays allocate the required number of rows, that is, the number of elements in the array at initialization time, and then store the pointer to the first record in the primary data structure table, and any index operation on the array is performed through pointer arithmetic, with bounds-checking.
Key-indexed maps create a secondary table, an additional cuckoo map with a given key type, and a record pointer as a value type. 
The physical pointer to the cuckoo map is also stored directly in the corresponding column in the main data structure.
The attribute columns in the main data structure hold the pointer to both collection types; however, accessing the base pointers bypasses the CC protocol as they are runtime constants.
Similarly, \systemname{} also generates a default destructor for each data structure that frees all memory allocated for the corresponding instance.

When a method accesses an attribute backed by the storage layer, the generated code of the method passes the physical record reference and the transaction object.
Then, the storage layer extracts the table ID and record offset and then executes the requested operation on the backing table. 
For each operation with side effects, create, update, or delete operations, the storage layer also inserts the change-log in the transaction's log to support rollback in case of aborts.


%
\section{Evaluation}
\label{sec:eval}




This section includes the results of our experimental evaluation.
First, we describe the hardware that we used to execute our experiments, and then some essential details of our software. Finally, we present our analysis, categorizing them with a focus on each aspect of the proposed approach, including composability, scalability, and adaptivity of declaratively generating CDS.

\subsection{Hardware \& Software setup} \label{sec:eval_setup}
\noindent\textbf{Hardware.}
All the experiments were conducted on a server equipped with
2x12-core Intel Xeon Gold 5118 processor clocked at 2.30 GHz, with Hyper-Threading on there is a total total of 48 logical cores and 384~GB of DRAM.
For all scalability experiments, we pin each thread to all first physical cores incrementally, then hyper-threads, before pinning threads on the next CPU NUMA node.
%


\noindent\textbf{Software.}
We implemented the DCDS framework in \systemname{}. 
\systemname{} uses oneAPI's thread-building blocks (TBB)~\cite{githubGitHubOneapisrconeTBB} library for memory allocations and read-write locks, and supply hints to use huge pages, if available. For code-generation and just-in-time compilation, \systemname{} employs LLVM compiler infrastructure (version 14). \systemname{} uses cuckoo-hashing~\cite{DBLP:conf/eurosys/LiAKF14} for key-value indexed items, and for the scope of this paper, we use JIT code-generation for all experiments, that is, all data structures are generated, compiled and executed, at runtime, in the same process.


\subsection{Usecase Specialization}

\fixme{idea: maybe write that very extreme example of a list converging to a single integer in the intro or manifesto, we don't do, but ideally, that should be automatically done. that would be a cool example.}


In this section, we show the applicability and impact of use-case-driven specialization. 
Data structure specifications are intended to be general and reusable without imposing a cost for generality. 
Consider a doubly linked list declared in \systemname{}. 
\cref{algo:list} shows the high-level pseudocode of two methods, pop\_front, and push\_back, while listing signatures of some of many the many methods that may be provided by a generic doubly linked list.

Consider a user who wants to use the declared doubly linked list as a component of their own data structure, named MyCDS.
The user will declare and compose the serial specification of the MyCDS, and then, at build time, the logical optimizer of \systemname{} will optimize and specialize the composed structure based on the declared requirements.
In the case where the doubly linked list is to be used as a first-in-first-out (FIFO) queue manner only, the logical optimizer will detect and remove any unused functionality from the composed list, keeping only the pop\_front and push\_back functions in scope.
With the remaining methods, the $prev$ attribute of the $Node$ type is now a write-only attribute and thus cannot alter the functionality of the methods. \systemname{} will therefore remove the $prev$ attribute and the corresponding statements that write to it on lines 7 and 14 of \cref{algo:list}.
This reduces the amount of work that is performed in the methods that are used by MyCDS.. 


\begin{figure}[t]
    \centering
\begin{lstlisting}[language=c++, firstline=0]
LinkedList {
    bool LL::pop_front(value_t &val){
        if(!head) { reutrn false; }
        if(head == tail) { tail = nullptr; }
        tmp = head;
        head = head->next;
        head->prev = nullptr; // prev is WO
        val = tmp->value;
        return true;
    }
    
    void LL::push_back(value_t value){
        node = Node(value);
        node->prev = tail;  // prev is WO
        if(!tail){ tail = head = node;} 
        else { tail->next = node; } 
        tail = node;   
    }
    
    // Unused functionality in the case when doubly linked list is being used in a FIFO manner:
    void LL::push_front(value_t value);
    bool LL::pop_back(value_t &val);
    bool LL::empty();
    // more unusued functions...
}

MyCDS{
    LinkedList list;
    
    void push(val){ list.push_back(val) }
    value_t pop(val){ return list.pop_front(); }
}
\end{lstlisting}
    \caption{Usecase-driven specialization of doubly- to singly-linked list for FIFO usages}
    \label{algo:list}
\end{figure}

\begin{figure}[ht]
    \centering
    \includegraphics[width=\columnwidth]{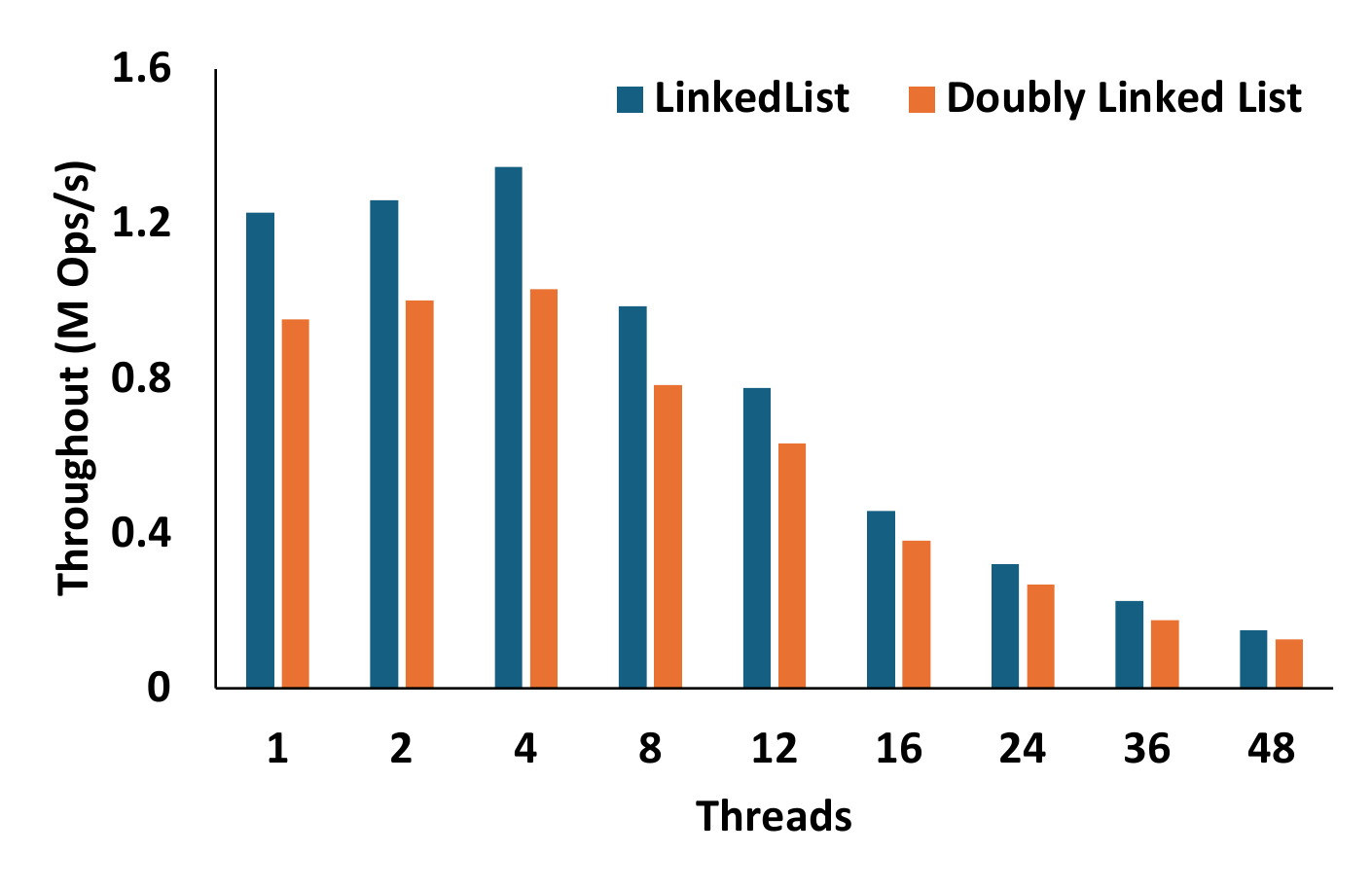}
    \caption{Specializing doubly linked list to FIFO use case}
    \label{fig:eval_fifo_ll_dll}
\end{figure}

\noindent\textbf{Workload.}
We experimentally evaluate and compare the performance of a doubly linked list compared to the case specialized singly linked list, and report the performance scalability in \cref{fig:eval_fifo_ll_dll}. We scale threads from 1 to 48, where each thread performs 100,000 operations in a closed loop. In each iteration, the thread either pushes a new value to the back of the list or pops from the front, with equal probability. 

\noindent\textbf{Baseline.}
For a fair comparison, we compare the performance scalability of the automatically optimized FIFO list in \systemname{} with a doubly linked list, also declared in the \systemname{}. The optimized list is derived from the base doubly linked list by the logical optimizer.

\noindent\textbf{Scalability.}
\cref{fig:eval_fifo_ll_dll} shows the result of our experimental evaluation.
Overall, the performance of both lists degrades with an increased number of threads as all operations are bottlenecked on either pushing or popping from the tail or the head, respectively. Further, a singly linked list performs better as it does not have to perform extraneous synchronization and updates on the $prev$ attribute of the doubly linked list, converging the data structure to a single linked list automatically given the declared use case.

\subsection{Composable Concurrency}

In this section, we evaluate the efficient composability of data structures generated by \systemname{}. We show that a simple LRU (least-recently-used) container generated with \systemname{} achieves better scalability then a implementation from a general purpose data structure library. In what follows, we first explain the functionality of a LRU container, then explain the workload used in the scalability evaluation, and then present the results of our experimental evaluation.

\begin{figure}[ht]
    \centering
    \includegraphics[width=0.8\columnwidth]{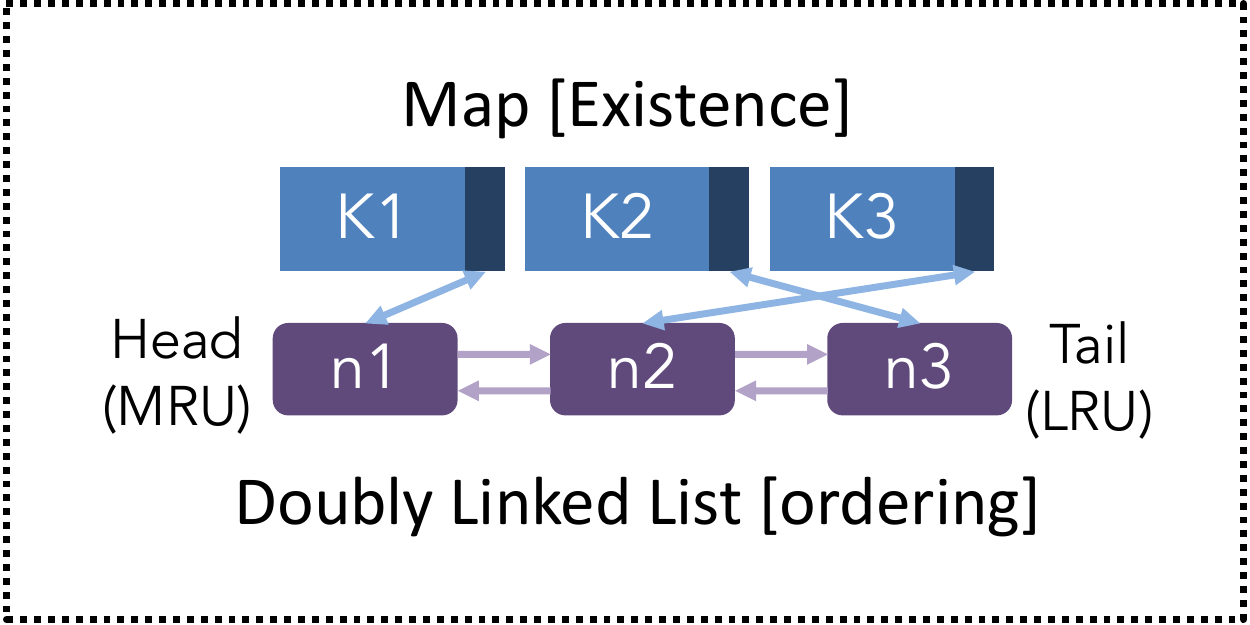}
    \caption{Composite Data Structure: LRU Container with a map and a doubly linked list}
    \label{fig:lru_list}
\end{figure}

\noindent\textbf{The LRU Container Data Structure.}
A LRU container is a mapping of keys to values with a fixed capacity (N) that maintains an ordering on the keys based on the most recent operation on each key. This enables identifying the most- (MRU) or least recently used (LRU) items.
A naive LRU is implemented through a doubly-linked list, where each $insert(key, value)$ operation inserts the key-value pair if it does not exist or moves the existing key to the head (MRU end) of the list. If the list size grows beyond the maximum capacity, a key-value pair is evicted from the tail (LRU).
In naive LRU, the existence check for a key requires traversing the list, asymptotically costing O(N).
An optimized implementation, shown in \cref{fig:lru_list}, utilizes a map in addition to the doubly linked list for existence check, reducing the asymptotic search cost from O(N) to O(1).
In a concurrent LRU container, the access operation requires atomicity across the operations on the map and the doubly linked list; for example, in the case of eviction, the removal from the map, as well as removal from the list, which includes unlinking and update the tail (LRU) end of the list.

In \systemname{}, the LRU container composes a doubly linked list, which is declared separately in \systemname{} with a key-indexed map. 
The nodes of the doubly linked list store a key, a value, and forward and backward pointers
The keys of the map are the same as in the LRU container interface, and the values of the map are pointers to the corresponding node.
The declared LRU container exposes an insert method and a find method. 
The insert method takes key and value arguments that return true if the key was inserted or false if the key already exists.
In either case, the node of the list containing the key is placed/moved to the head of the list.
If the container is at capacity, it will evict the least recently used key from the tail of the list. 
The find method takes a key and a pointer to a key-type variable. It returns true if the key was found, setting the variable pointed to to the found key while moving the corresponding node to the head of the list, and returns false if the key was not found.

\noindent\textbf{Workload.}
We evaluate LRU containers using a key domain of $0-1M (2^{20})$ and a maximum capacity of the containers of $1024 (2^{10})$. We scale the number of threads from 1 to 48, using the physical then hyper threads of the first CPU NUMA node before scheduling the threads on the second CPU NUMA node. We run the benchmark in two configurations: with keys drawn from the domain with a uniform random distribution and with a Zipfian distribution with a theta of 0.4. In each benchmark run, each thread executes one hundred thousand insert operations.


\noindent\textbf{Baseline.}
We compare \systemname{} with an open-source LRU container that is a part of oneAPI TBB (tbb::concurrent\_lru\_cache)~\cite{githubGitHubOneapisrconeTBB}.
TBB's LRU implementation provides a single accessor function that takes a key as input and returns a reference counted handle. 
The container stores all items in use (items with a reference count greater than 0), plus a number of unused items (items with a reference count of 0). 
The container only evicts unused items. 
When the count of unused items exceeds the capacity, the least recently used unused item is evicted. 
This essentially allows the list to grow beyond the specified capacity as long as each of the keys has a reference held external to the container.
In our setup, the experimental loop of each thread accesses a key but does not hold the reference. So, on each access iteration, only a single object is used and referenced from the list, preventing the container from exceeding the specified capacity.


\noindent\textbf{Scalability.}
The TBB implementation internally uses delegation-based concurrency; a single queue is maintained for insertion and eviction operations, protected by a single lock.
Operations from all threads are placed into the queue and one thread executes the operations. 

It uses a c++ std::map to check if the key is already present in the LRU list, the internal operations on the std::map are serialized through the operation queue. 
Regardless of uniform or zipfian access, the throughput of TBB's LRU list drops after two threads due to the contention on the operation queue object.
An LRU container has an inherent bottleneck on the head of the list, as on each operation, the head will be updated. However, in \systemname{} the granularized CC scales better than TBB as the doubly linked list provides concurrent accesses to the head and any other node in the list, while also interleaving concurrent operations across the map and list, unlike TBB where the internal operations are run in serial on a single thread through the operation queue. 

\fixme{Summary still valid?}
\systemname{} enables efficient composability by interleaving concurrent operations while eliminating the need of a serial synchronization wrapper across composed structures.

\begin{figure}[ht]
  \centering
  \begin{minipage}[t]{0.95\columnwidth}
  \centering
  \subfigure[Uniform Access]{\centering\includegraphics[width=\textwidth]{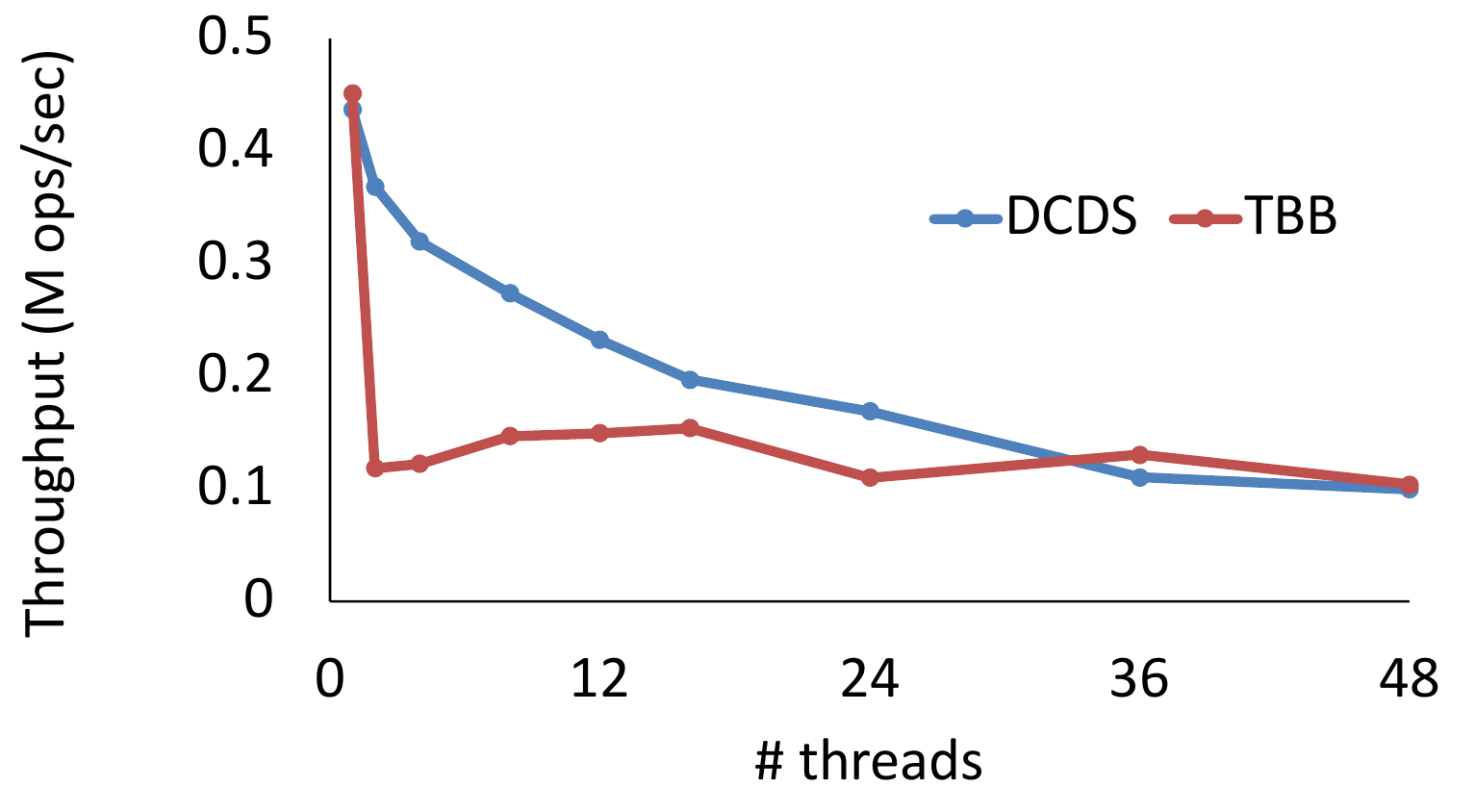} 
    \label{fig:eval_lru_uniform}
  }%
  \end{minipage}%
  \hfill
  \begin{minipage}[t]{0.95\columnwidth}
  \centering
    \subfigure[Zipfian (0.4) Access]{\centering\includegraphics[width=\textwidth]{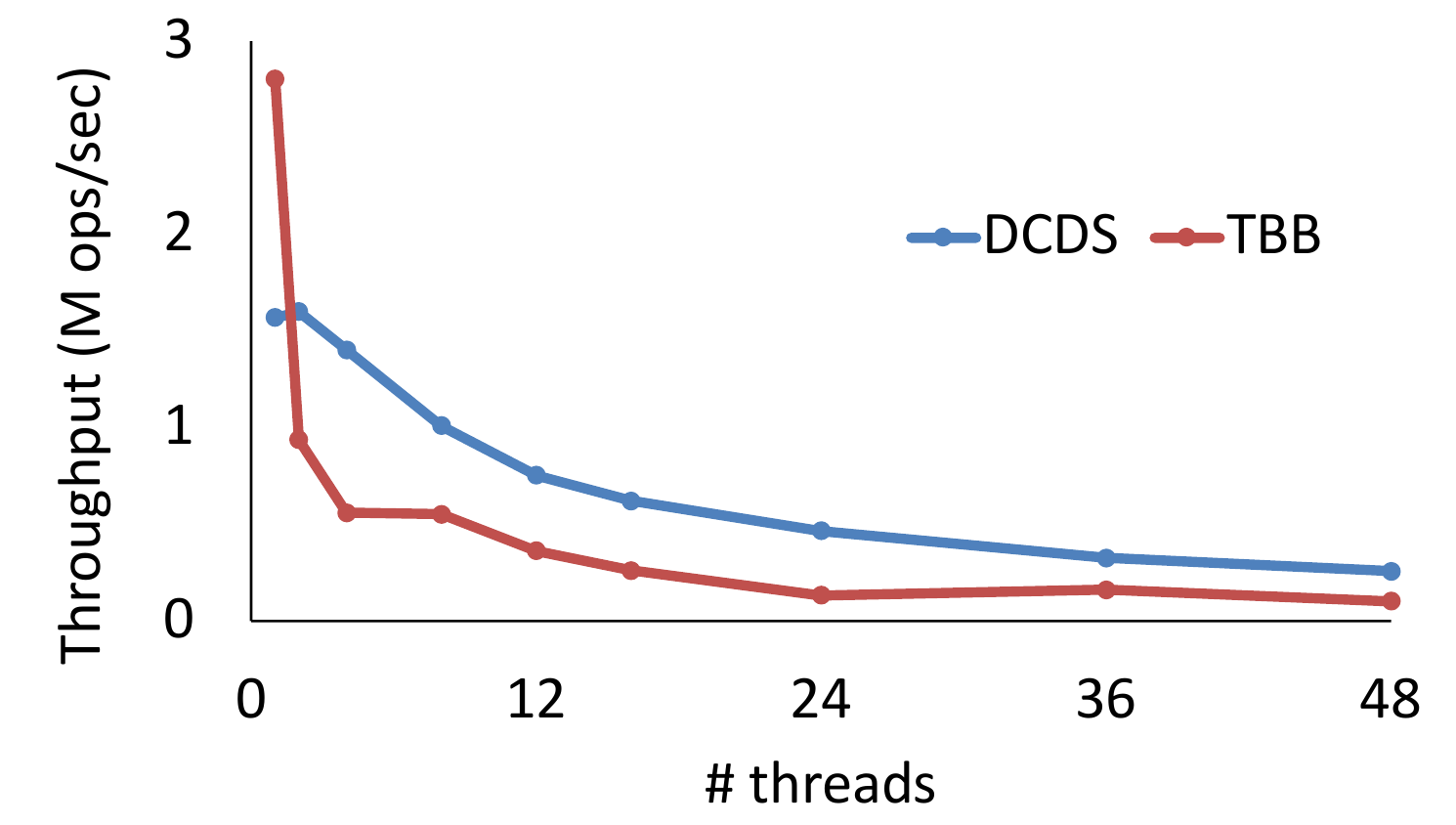} 
    \label{fig:eval_lru_zipf4}
    }%
  \end{minipage}%
  \hfill
  \caption
    []{%
      Scalability of a LRU container generated by \systemname{} compared to oneAPI TBB (tbb::concurrent\_lru\_cache)
      \label{fig:eval_lru}%
    }%
\end{figure}


\subsection{In-process OLTP}
In this section, we evaluate the general applicability of \systemname{} for standard transactional workloads. We show that using \systemname{} for simple transaction workload outperforms state-of-the-art in-memory DBMS. This is due to the fact that 1) \systemname{} generates and specializes concurrency control to the pre-defined workload operations, and 2) \systemname{} does not incur the generalization overhead of a DBMS such as multi-versioning, pooling, transaction queues, etc.

\noindent\textbf{Workload.}
We performed experimental evaluation using YCSB~\cite{DBLP:conf/cloud/CooperSTRS10} benchmark. YCSB is a key-value style benchmark designed for testing and analyzing the scalability of transactional workload in a system. We implement YCSB-style workload in \systemname{} by declaring a data structure having a fixed-size array of $YCSB\_ITEM$ type. $YCSB\_ITEM$ contains one to ten 64-bit integer columns, depending on the specific experiment. We set the total number of records as $1M * num\_workers$ to have similar record access and write distribution across workers. To simulate frequent data structure operations, we read or update one record, selected either uniformly or through Zipfian distribution, with different read and write ratios. 

\noindent\textbf{Baseline.} 
We compare \systemname{} with Proteus, which is a state-of-the-art in-memory OLTP DBMS~\cite{DBLP:journals/pacmmod/RazaCAA23}~\cite{DBLP:conf/sigmod/RazaCAA20}. Proteus employs MV2PL concurrency control with snapshot isolation and maintains a primary hash-based index per table, implemented using cuckoo hashing~\cite{DBLP:conf/eurosys/LiAKF14}. The index contains the transactional timestamps, version pointer, and logical record pointer in the form of RowIDs (RID). 
The storage layout for OLTP is columnar.
Proteus allocates a warm pool of huge pages at the system startup and then allocates any required memory from the pre-allocated memory.
Workers are started and pinned first on each physical thread and then hyper-thread within a socket before allocating from the other CPU NUMA node. All transaction workers generate and execute transactions in a closed loop, simulating a full transactional queue.

\begin{figure}[ht]
    \centering
    \includegraphics[width=\columnwidth]{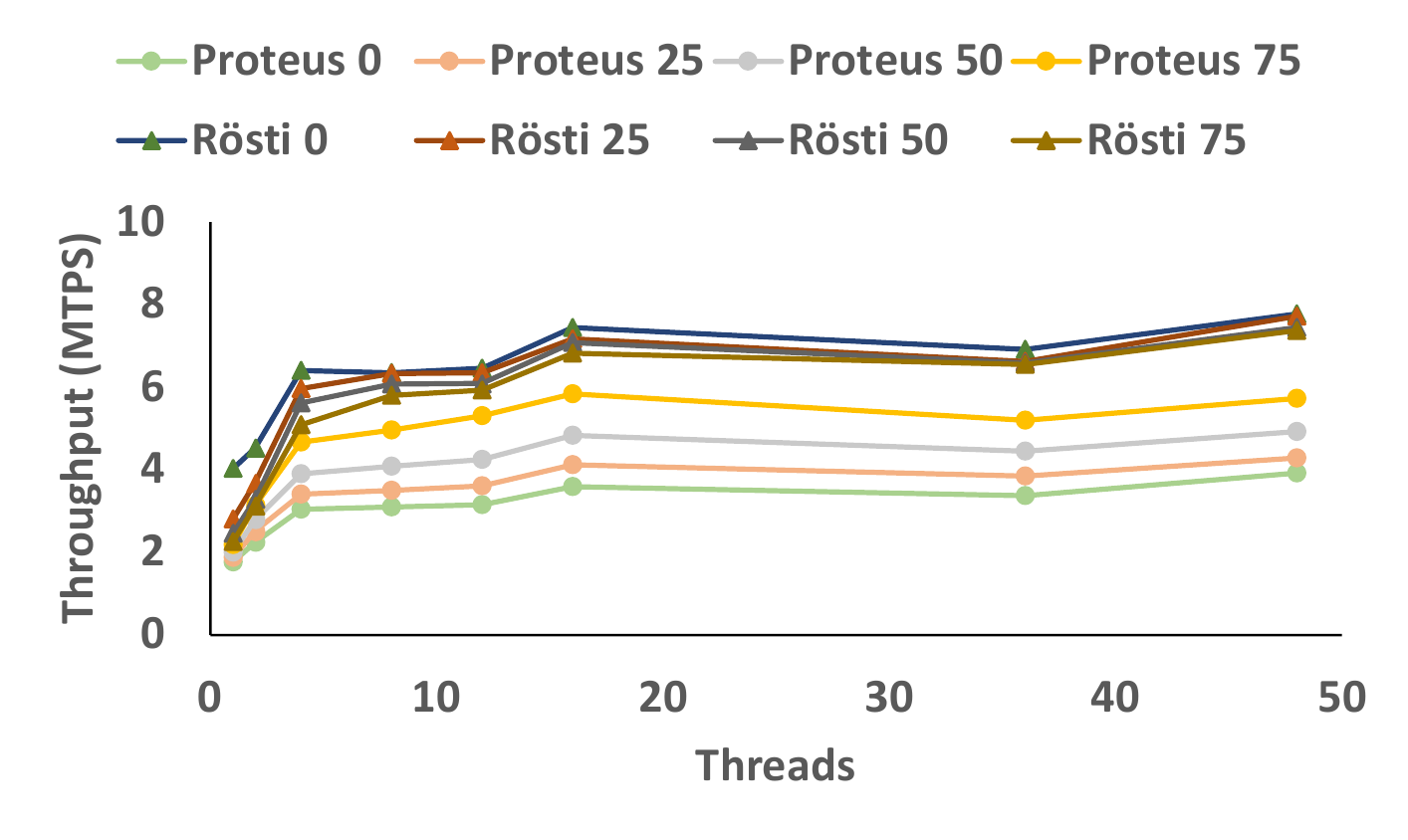}
    \caption{YCSB scalability compared to in-memory DBMS}
    \label{fig:eval_ycsb_thread_scaling}
\end{figure}
\noindent\textbf{Scalability.}
\cref{fig:eval_ycsb_thread_scaling} evaluates and compares \systemname{}'s scalability with Proteus, using the YCSB benchmark. In this experiment, we use uniform distribution for record selection and then analyze both systems' scalability through varying read-write ratios while increasing the number of workers from 1 to 48. Do note that threads 1-12 are the first physical cores, then 13-24 threads are hyper-threads collocated in the same socket and, afterward, in the second CPU NUMA socket. Then, we report the throughput of each system in million transactions per second (MTPS) for each setup.
Both systems scale until 4-threads, where it starts to saturate the micro-architectural resources, including but not limited to store queues, caches, and random read-write performance. \systemname{} has an overall higher throughput than Proteus due to an overall less amount of work. Proteus has to perform multi-versioning for updates, while for both reads and writes, it employs a spin-latch over the record during the record or version access, whereas \systemname{} uses simple reader-writer locks. In summary, \systemname{} outperforms in-memory DBMS due to the simple fact that it specializes concurrency control to the target workload, avoiding unnecessary synchronization which a DBMS couldn't, and reduces the total amount of work given the specialized generated data structure.

\section{Related Work}
\label{sec:related}
Transactional Memory (TM) emerged as a promising approach to simplify concurrent programming through atomic multi-word compare and swap, initially proposed as a hardware mechanism~\cite{DBLP:conf/isca/HerlihyM93} and then in software (STM) using existing hardware synchronization primitives~\cite{DBLP:journals/dc/ShavitT97}. 
At its core, STM is an abstraction for concurrency control, and many concurrency control schemes have been created and used to implement STM~\cite{DBLP:conf/wdag/DiceSS06,DBLP:conf/pldi/DragojevicGK09,DBLP:journals/tpds/FelberFMR10,DBLP:conf/ppopp/MaratheM07,DBLP:conf/wdag/SpearMSS06}. 
There are multiple interfaces to use STM. 

Word-based STM libraries, such as TL2~\cite{DBLP:conf/wdag/DiceSS06} and SwissTM~\cite{DBLP:conf/pldi/DragojevicGK09}, work by intercepting and synchronizing direct memory accesses. 
Word-based synchronization is intended for low-level unmanaged languages, such as C and C++, that allow direct access to memory without a high-level language that mediates access to memory.
The lack of run-time code generation and reflection in such environments requires the user of a word-based STM library to determine which memory accesses need to be transactional, and for each call STM library methods to perform the memory access. 
With compiler support for an unmanaged language, a user can write code and annotate blocks that must be performed atomically; the compiler then instruments memory accesses in the annotated blocks to intercept and synchronize memory accesses~\cite{archiveIntelCompiler,schindewolf_towards_2009,DBLP:conf/wscad/HonorioCB18,archiveAlphaWorksCC,DBLP:conf/cgo/WangCWSA07}. 
However, these approaches suffer from a lack of application-level knowledge and can only conservatively apply optimizations.
This causes the compiler to generate excessive read/write barriers, resulting in higher overheads and false conflicts between transactions~\cite{DBLP:conf/spaa/YooNWSAL08}. To address over-instrumentation, some compiler-assisted STM implementations have a mechanism for the programmer to convey application-level knowledge to the compiler, such as which data is private and does not need to be transactionally accessed~\cite{DBLP:conf/spaa/YooNWSAL08,DBLP:journals/cacm/CascavalBMCWCC08}. This places the onus on the programmer to understand the performance of their transactional code in order to optimize it and also risks incorrectness if the programmer accidentally annotates data that does not need to be transactionally accessed. 

Object-based STM intercepts and synchronizes calls to object methods, leveraging the language features of object-oriented languages.
The approaches taken by Herlighy et al. and Fraser
make shadow copies of each transactional accessed object ~\cite{DBLP:conf/podc/HerlihyLMS03,DBLP:conf/lcr/MaratheSS04,DBLP:phd/ethos/Fraser04}.
Herman et al. introduce Software Transactional Objects STO, which delegates all concrete
locking, version verification, and data structure modification to the data structures themselves, while the core STO system implements a commit protocol leveraging the API that the transactional objects must implement~\cite{DBLP:conf/eurosys/HermanIHTKLS16}.
This enables STO transactional data structures to leverage their semantics to implement specialized concurrency control mechanisms at the cost of increased complexity to implement a transactional data structure.  
Spiegelman et. al proposed a method to compose transactional data structures as long as the base transactional data structures implement a concurrency control protocol that meets the semantics of their API~\cite{DBLP:conf/pldi/SpiegelmanGK16}. This approach allows for the straightforward composition of existing transactional data structures. However, implementing a new base transactional data structure is more involved than implementing a lock-free data structure from scratch.

\section{Outlook and Conclusion}
\label{sec:conclusion}
\fixme{Consistency check on plural acronyms across paper, e.g use DBMSs }





In this paper, we have presented an innovative approach to designing and implementing concurrent data structures (CDS) through the Declarative Concurrent Data Structures (DCDS) framework. DCDS facilitates the declaration of data structure attributes and methods, abstracting away the intricate details and complexities of the concurrency control mechanisms. Our approach significantly eases the CDS development process by enabling developers to focus on the functional requirements of their data structures, while the DCDS framework generates concurrent and scalable code.

We demonstrate the applicability of DCDS through our prototype system \systemname{}.
\systemname{} generates scalable data structures, achieving up to 2x speed up relative to a state-of-the-art in-memory database, and maintains this scalability on composed data structures, demonstrating up to 2x speed up of LRU lists compared to a specialized LRU cache implementation.

The DCDS framework enables future research directions beyond static in-memory CDSs. 
The most appropriate concurrency control algorithm depends on workload, such as optimistic CC for low-contention while pessimistic for high-contention workloads~\cite{DBLP:journals/pvldb/WuALXP17, DBLP:journals/pvldb/AppuswamyAPIA17}. Additionally, accurate workload-driven optimizations can only be done at runtime. 
The DCDS framework will enable dynamic tracing of workload at runtime and then adapt the optimizations and CC mechanism accordingly.
Further, DCDS abstracts the physical storage from the specification of the data structure. 
The implementation has the flexibility to alter the layout, for example, to efficiently use CPU caches, as well as where the data is stored. 
This means that a DCDS implementation could be used to generate data structures that can spill from memory to persistent storage, or generate data structures for tiered memory systems.


\bibliographystyle{ACM-Reference-Format}
\bibliography{references.bib,hamish_zotero.bib}

\end{document}